# An Easy-to-Use-and-Deploy Grid Computing Framework

Gaurav Menghani, Anil Harwani, Yash Londhe, Kalpesh Kagresha
*Department of Computer Engineering,*
*Thadomal Shahani Engineering College, Mumbai – INDIA*
gaurav.menghani@gmail.com, harwanianil@gmail.com, yashlondhe@gmail.com, kkagresha@yahoo.com

**Abstract**

*A few grid-computing tools are available for public use. However, such systems are usually quite complex and require several man-months to set up. In case the user wishes to set-up an ad-hoc grid in a small span of time, such tools cannot be used. Moreover, the complex services they provide, like, reliable file transfer, extra layers of security etc., act as an overhead to performance in case the network is small and reliable.*

*In this paper we describe the structure of our grid-computing framework, which can be implemented and used, easily on a moderate sized network.*

## 1. Introduction

While most grid toolkits do an excellent job of managing a large-scale grid, installing them in the first place is time-consuming, not to mention complicated. [1]. For large problems, which might require several CPU years of time, and need the collaboration of thousands of nodes, which might lie outside the administrative domain of the user, the use of such toolkits is justified.

However, if we have a smaller problem (of the order of a couple of CPU days of time), then the overhead of data-transfer to-and-from the systems outside the network becomes significant. Ideally, if the organization has sufficient resources within its boundaries connected through a reliable network, this can be avoided. Also, the additional layers of reliable file-transfer, security etc. can be done away with. What we aimed was to design a high-performing, low-overhead grid computing framework for small networks, which could be setup with ease.

The framework is also especially useful for those who do not want to put up a grid for permanent use. In such cases, investing time and money into learning how complex grid computing tools work, and deploying them, is not an efficient approach. The framework can also be used to set up a cluster out of shared machines with standard hardware.

## 2. Objectives

The following activities need to be performed, and in order:

1. Accept the problem to be solved from the user, consisting of parallel code units called Tasks, dependency matrix of tasks, and other vital data.
2. Distribute these tasks while taking in consideration the inter-dependency of tasks, and using a grid-scheduling algorithm.
3. Solve tasks at worker nodes; record the output and errors (if any).
4. Worker nodes send the output (errors if any) and, performance logs to the supervisor node.
5. Collect outputs and logs from workers.
6. Update worker performance statistics.
7. Ability to cancel / re-do the execution of tasks dynamically at 'checkpoints', hence modifying the execution on the fly. This is especially handy when the output of one task can possibly alter the execution of other tasks, for example when the grid is being used for breaking an encryption key.
8. Arrange outputs as desired by the user and present it to the user.

## 3. Methodology

The framework primarily involves three entities.
1. User
2. Supervisor Node
3. Worker Node(s)

The user is concerned with solving a problem, which is composed of a set of independent / interdependent tasks with separate input sets. The user submits the complete problem to the supervisor node.

The problem consists of the following:

**1. Problem Solving Schema (PSS)**: It is an XML document which describes the problem by giving a short summary of the problem, the name of the task files, priority of the tasks, a dependency matrix of the tasks etc.

The XML tags are used to describe the essential parts of the problem. Another utility of this way of input is that a variety of interfaces can be built.

**2. Task File(s):** These file(s) are the programs that the user wants to be executed at the worker nodes. The user provides commands for compilation and execution, in the PSS.

Hence, the programs can be originally written in any language, or can be the combination of modules written in different languages. The PSS extends the flexibility to the user.

**3. Task File Input Sets**: These sets are auxiliary files that accompany the task file(s) and are used as input by them.

The user is responsible for breaking the original sequential program into independent task files and the task file input sets. This step is analogous to the Map function of the Map-Reduce paradigm [2].

**4. Result Compilation Program (RCP)**: This program runs at the supervisor and processes all the output generated by the independent task files when they are remotely executed on worker nodes.

This step is analogous to the Reduce function of the Map-Reduce paradigm [2].

**5. Execution Monitor Program (EMP)**: This program is executed after certain 'checkpoints' (which the user specifies in the PSS). Here, the Supervisor stops issuing new tasks to workers, executes the EMP as per the execution commands in the PSS, and waits for the EMP's commands. The EMP can either ask the Supervisor to stop all or specific tasks, or to redo all or specific tasks, or continue distributing tasks normally.

The above set of files is collectively known as the *Problem*. Upon receiving the problem, the supervisor first parses the PSS. It then uses the dependency matrix to topologically sort the tasks to be performed and then queues as per their priority. The supervisor then distributes the tasks amongst the workers using a specialized Grid Scheduling Algorithm (GSA), which intends to maintain the economy of resource consumption as well as the speed of task solving. As explained, the supervisor executes the EMP as and when the checkpoint tasks are completed.

The supervisor provides the worker with the following:

**1. Task File**: This is one of the tasks in the problem.
**2. Task Input File(s) and other Auxiliary Files**
**3. Task Compilation Commands**
**4. Task Execution Commands**
**5. Task Priority**: This value is obtained from the PSS and is used to set the 'nice' value of the program executable.
**6. Task Timeout**: This is the time limit for the execution of the task.

This collection of information is known as a *Task*. The supervisor packs these files into a single gzip tar archive and sends it to the respective worker nodes. The worker on receiving the archive, unzips it and compiles the task file. The compiled executable is executed on the worker machine with the specified priority up till either the task completes execution (successfully or unsuccessfully) or the task times out.

Whatever be the result, the supervisor is obligated to send the following to the user:

**1. Task Output**: The output produced by the task.
**2. Error Log**: Any errors (compile time or run time)

**3. Task Statistics**: Time taken for the execution of the task.

This packet of information is collectively known as the *Task Execution Result*.

The supervisor continuously keeps receiving the Task Execution Results. However, some of them might be solved and some might have generated errors. If the task has generated errors, it might be compile-time or run-time errors.

But since, a worker node on the grid does not guarantee a perfect environment always, the supervisor retries sending the same Task to other client nodes. Repeated failure to get a valid Task Execution Result leaves the supervisor with no choice, but to abandon the task as well as the problem. The
supervisor records the errors in an error log.

However, if the supervisor receives a valid Task Execution Result, it stores the result and continues with its work. If the completed task was a checkpoint, the EMP is executed and the supervisor waits for its commands, and takes actions accordingly. When all the tasks have been completed, the supervisor then executes the Result Compilation Program to collate all the output as per the user.

The problem is said to be solved successfully only after the RCP has produced the final output.

Regardless of whether the problem was solved successfully or not, the supervisor is required to present the following information to the user:

**1. Problem Output**: The output generated by the RCP. None, if the problem was abandoned.

**2. Task Execution Result(s):** All the Task Execution Result(s) are stored here so that the user can debug and check the outputs.

**3. Problem Statistics**: These are statistics pertinent to the execution of the problem, like total computing resources consumed etc

## 4. Implementation

Both the supervisor and worker programs have four main threads running concurrently.
1. Ping thread
2. Ping listener thread
3. Coordinator thread
4. File transfer thread

Ping and ping listener threads are used so that the Supervisor and Workers know that the other is active. While the supervisor continuously sends a ping containing its address to all the machines on the network, the workers use the ping listener thread to know the supervisor's address.

The worker pings the supervisor on the address it received. For scheduling purposes, each worker also runs a special thread, which evaluates its computational load and network latency up to that machine. These values are known as the performance and network metrics respectively, and are piggybacked with the ping to the supervisor. The metrics help the supervisor in dynamically altering the task scheduling decisions (scheduling is explained in detail in section 5).

The file transfer thread, as the name suggests is used to transfer files between machines.

The coordinator thread of the supervisor is the heart of the framework. It initiates the parsing of the PSS, prepares the task archives, initiating the file transfer thread, sending commands to the worker nodes etc.

Similarly the coordinator thread of the worker receives commands from the supervisor to which it responds by receiving files using the file transfer thread, unzipping the task received, compiling and executing the task files using the commands provided, and sending the output back using the file transfer thread again.

## 5. Scheduling Algorithm

A grid is composed of a large number of heterogeneous resources. The workers might have different computational potentials, and the latency between the supervisor–worker link might vary. A slow worker connected through a high-speed link might be a better option than a fast worker connected through a very-slow link. Hence, two metrics were designed to gauge the computational capacity of the workers, and the latency of the supervisor-worker link. The first metric is the performance metric, and is roughly proportional to the number of floating point calculations that can be done by the worker node, which is found out by operations like matrix multiplication and inversion. The second metric is the network metric, and is the expected latency between the supervisor and the worker calculated by using its statistical variation in the past.

The metrics can be used as an input to the grid scheduling algorithms. These algorithms are about finding an optimal mapping of tasks to machines. However, it has been shown that the problem of finding such a mapping is NP-Complete [3]. Several heuristics have been proposed by researchers, which try to optimize the mapping of tasks [4, 3], including several nature-based heuristics [4,5,6].

For the purpose of choosing a suitable algorithm for the Framework, we studied the heuristics mentioned previously. However, MCT, Min-Min, Sufferage, Genetic Algorithms and Simulated Annealing Algorithms were found to be efficient and hence were tested further.

The Genetic Algorithm implementation by Braun et al. [1], was found to be one of the best heuristics in all type of ETC matrices. However, the results and especially the execution time can be improved with some modifications.

Research in [7] led us to a modified Genetic Algorithm. This algorithm is different from [1], in multiple respects. It depends on the Segmented Sympathy heuristic.

The Segmented Sympathy heuristic makes use of the sympathy metric, which is defined as $s_i = E(c_i) \times V(c_i)$. Where $E(c_i)$ is the mean completion time of task $i$ on all machines, and $V(c_i)$ is the variance of the completion times of task $i$ on all machines. The sympathy metric gives a rough idea of the improvement possible in the makespan, if that particular task is assigned to a machine where it will achieve the best completion time.

---

**Segmented Sympathy**

(1) **start**
(2) Sort each task by their respective sympathy metrics.
(3) Partition the tasks evenly into N segments.
(4) Starting with the segment with the highest values of sympathy metric, apply Min-Min individually on the segments.
(5) **end**

---

The modified Genetic Algorithm in [7] uses the mappings generated by Min-Min, Segmented Min-Min [8] and Segmented Sympathy as the seeds to the algorithm. It also differs significantly with respect to the elitism criteria, mutation and crossover operators, population size, etc.

Simulation results in [7] show that it is about 160% faster than the Genetic Algorithm provided by Braun et. al. Makespans were better by 3.42% in an average case, and up to 8.34% in the best case . The algorithm was chosen for scheduling tasks in the framework.

# 6. Advantages of the Framework

The framework offers the following advantages:
1. It is extremely easy to use. The user needs to know only the PSS XML tags to use it. It is easy to design a user-friendly web-based or visual interface for the user, which can simply convert the user's inputs to the PSS format.
2. It is easy to deploy. The command-line version of the framework has no dependencies. On a standard Linux machine, it runs out-of-the-box.
3. It is lightweight and has low memory requirements.
4. The user can use any language for solving the problem, and the framework isn't tied down with any particular language.
5. The framework allows changing the flow of task-execution dynamically with the help of the Execution Monitor Program.

# 7. Further Work

Though we successfully tested the framework in a controlled moderate sized local network, which was our target environment, to empower the user to use the framework in a completely unmonitored large scale network, some additions need to be made.

These include the following,
1. Implement a 'Grid Manager' level, which can manage multiple supervisors, each of which can instead distribute tasks in a network.
2. Add authentication for an unsecured network.
3. Allow multiple supervisors to function in harmony on the same subnet.
4. Add capability to handle extreme load.